\documentclass{webofc}
\usepackage[varg]{txfonts}   
\wocname{QUARKS-2016}
\woctitle{19th International Seminar on High Energy Physics}
\begin{document}
\title{$E_6$ inspired composite Higgs model and $750\,\mbox{GeV}$ diphoton\\ excess}
%
%

\author{\firstname{Roman} \lastname{Nevzorov}\inst{1,2}\fnsep\thanks{\email{roman.nevzorov@adelaide.edu.au}} \and
        \firstname{Anthony} \lastname{Thomas}\inst{1}
}

\institute{ARC Centre of Excellence for Particle Physics at the Terascale and CSSM,
Department of Physics, The University of Adelaide, Adelaide SA 5005, Australia
\and
        Institute for Theoretical and Experimental Physics, Moscow 117218, Russia
          }

\abstract{%
In the $E_6$ inspired composite Higgs model (E$_6$CHM) the strongly interacting sector
possesses an SU(6)$\times U(1)_B\times U(1)_L$ global symmetry. Near scale $f\gtrsim 10\,\mbox{TeV}$
the SU(6) symmetry is broken down to its SU(5) subgroup, that involves the standard model (SM)
gauge group. This breakdown of SU(6) leads to a set of pseudo--Nambu--Goldstone bosons (pNGBs)
including a SM--like Higgs and a SM singlet pseudoscalar $A$.  Because of the interactions between
$A$ and exotic fermions, which ensures the approximate unification of the SM gauge couplings and
anomaly cancellation in this model, the couplings of the pseudoscalar $A$ to gauge bosons get induced.
As a result, the SM singlet pNGB state $A$ with mass around $750\,\mbox{GeV}$ may give rise to
sufficiently large cross section of $pp\to \gamma\gamma$ that can be identified with the recently observed
diphoton excess.}
\maketitle
\section{Introduction}
\label{intro}
The discovery of the Higgs boson with mass $m_h \simeq 125\,\mbox{GeV}$
allows one to estimate the values of parameters of the Higgs potential. In the standard model (SM)
the Higgs scalar potential is given by
\begin{equation}
V(H) = m^2_H\, H^{\dagger} H + \lambda\, (H^{\dagger} H)^2\,.
\label{quarks-1}
\end{equation}
The $125\,\mbox{GeV}$ Higgs mass corresponds to $m^2_H \approx -(90\, \mbox{GeV})^2$
and $\lambda\approx 0.13$. At the moment current data does not permit to distinguish whether
Higgs boson is an elementary particle or a composite state. Although the discovered Higgs boson
can be composed of more fundamental degrees of freedom, the rather small values of $|m^2_H|$ and
the Higgs quartic coupling $\lambda$ indicate that the Higgs field may emerge as a pseudo--Nambu--Goldstone
boson (pNGB) from the spontaneous breaking of an approximate global symmetry of some strongly
interacting sector.

The minimal composite Higgs model (MCHM) \cite{Agashe:2004rs} includes weakly--coupled
elementary and strongly coupled composite sectors (for a recent review, see \cite{Bellazzini:2014yua}).
The weakly--coupled elementary sector involves all SM fermions and gauge bosons. The strongly coupled sector gives
rise to a set of bound states that contains Higgs doublet and massive fields with the quantum numbers of all SM particles.
These fields are associated with the composite partners of the quarks, leptons and gauge bosons.
The elementary states couple to the composite operators of the strongly interacting sector leading to
mixing between these states and their composite partners. In this framework, which is called partial compositeness,
the couplings of the SM states to the composite Higgs are set by the fractions of the compositeness of these states.
The observed mass hierarchy in the quark and lepton sectors can be accommodated through
partial compositeness if the fractions of compositeness of the first and second generation fermions are quite small.
In this case the flavour-changing processes and the modifications of the $W$ and $Z$ couplings associated with the
light SM fermions are somewhat suppressed. At the same time, the top quark is so heavy that the right--handed
top quark $t^c$ should have sizeable fraction of compositeness.

The strongly interacting sector of the MCHM possesses global $\mbox{SO(5)}\times U(1)_X$ symmetry that contains
the $\mbox{SU(2)}_W\times U(1)_Y$ subgroup. Near the scale $f$ the $\mbox{SO(5)}$ symmetry is broken down to $\mbox{SO(4)}$ so that
the SM gauge group remains intact, resulting in four pNGB states which form the Higgs doublet. The custodial global
symmetry $\mbox{SU(2)}_{cust} \subset \mbox{SO(4)}$ allows one to protect the Peskin--Takeuchi $\hat{T}$ parameter against~ new~
physics~ contributions.~ Experimental~ limits~ on~ the~ parameter~ $S$ imply that $m_{\rho}=g_{\rho} f \gtrsim 2.5\,\mbox{TeV}$,
where $m_{\rho}$ is a scale associated with the masses of the set of spin-1 resonances and
$g_{\rho}$ is a coupling of these $\rho$--like vector resonances. This set of resonances, in particular, contains
composite partners of the SM gauge bosons. Even more stringent bounds on $f$ come from the non--observation
of flavor changing neutral currents (FCNCs). In the composite Higgs models, adequate suppression of
the non--diagonal flavour transitions can be obtained only if $f$ is larger than $10\,\mbox{TeV}$.
This bound on the scale $f$ can be considerably alleviated in the models with additional flavour symmetries $\mbox{FS}$.
In the models with $\mbox{FS}=U(2)^3=U(2)_{q}\times U(2)_u \times U(2)_d$ symmetry, the bounds that originate
from the Kaon and $B$ systems can be satisfied even for $m_{\rho}\sim 3\,\mbox{TeV}$. In these models the appropriate
suppression of the baryon number violating operators and the Majorana masses of the left--handed neutrino can be
achieved if global $U(1)_B$ and $U(1)_L$ symmetries, which ensure the conservation of the baryon and
lepton numbers to a very good approximation, are imposed. Thus the composite Higgs models under consideration
are based on
\begin{equation}
\mbox{SU(3)}_C\times \mbox{SO(5)}\times U(1)_X \times U(1)_B \times U(1)_L \times \mbox{FS}\,.
\label{quarks-2}
\end{equation}
The couplings of the elementary states to the strongly interacting sector explicitly break the $\mbox{SO(5)}$ global symmetry.
As a consequence, the pNGB Higgs potential arises from loops involving elementary states. This leads to the suppression
of the effective quartic Higgs coupling $\lambda$.
\section{$E_6$ inspired composite Higgs model}
\label{sec-1}
In the $E_6$ inspired composite Higgs model (E$_6$CHM) the Lagrangian of the strongly coupled sector is invariant under
the transformations of an $\mbox{SU(6)}\times U(1)_B\times U(1)_L$ global symmetry. The E$_6$CHM can be embedded into
$N=1$ supersymmetric (SUSY) orbifold Grand Unified Theories (GUTs) in six dimensions which are based on the
$E_6\times G_0$ gauge group \cite{e6chm}. (Different aspects of the $E_6$ inspired models with low-scale supersymmetry breaking
were recently considered in \cite{King:2005jy}-\cite{King:2016wep}.) Near~ some~ high~ energy~ scale,~ $M_X$,~ the~
$E_6\times G_0$~ gauge~ group~ is~ broken~ down~ to~ the~ $\mbox{SU(3)}_C\times \mbox{SU(2)}_W\times U(1)_Y \times G$
subgroup where $\mbox{SU(3)}_C\times \mbox{SU(2)}_W\times U(1)_Y$
is the SM gauge group. Gauge groups $G_0$ and $G$ are associated with the strongly interacting sector. Fields belonging to this
sector can be charged under both the $E_6$ and $G_0$ ($G$) gauge symmetries. The weakly--coupled sector includes
elementary states that participate in the $E_6$ interactions only. Due to the conservation of the $U(1)_B$ and $U(1)_L$
charges all elementary states with different baryon and/or lepton numbers are components of different bulk  $27$--plets,
whereas all other components of these $27$--plets have to acquire masses of the order of $M_X$. All fields from the strongly
interacting sector reside on the brane where $E_6$ symmetry is broken down to the $\mbox{SU(6)}$ that contains an
$\mbox{SU(3)}_C\times \mbox{SU(2)}_W\times U(1)_Y$ subgroup. As a result at high energies the Lagrangian of the composite sector respects
$\mbox{SU(6)}$ global symmetry. The SM gauge interactions violate this symmetry.  Nevertheless, $\mbox{SU(6)}$ can remain an approximate
global symmetry of the strongly coupled sector at low energies if the gauge couplings of this sector are considerably
larger than the SM ones.

As in most composite Higgs models, the global $\mbox{SU(6)}$ symmetry in the E$_6$CHM is expected to be broken below scale $f$.
Here we assume that it gets broken to $\mbox{SU(5)}$ subgroup, so that the SM gauge group is preserved.
Since E$_6$CHM does not possesses any extra custodial or flavour symmetry, the scale $f$ must be much larger than the
weak scale, i.e. $v\ll f$. In particular, the adequate suppression of the FCNCs requires $f\gtrsim 10\,\mbox{TeV}$.
The $\mbox{SU(6)}/\mbox{SU(5)}$ coset space includes eleven pNGB states that correspond to the broken generators $T^{\hat{a}}$
of $\mbox{SU(6)}$. These pNGB states can be parameterised by
\begin{equation}
\begin{array}{c}
\Omega^T = \Omega_0^T \Sigma^T = e^{i\frac{\phi_0}{\sqrt{15}f}}
\Biggl(C \phi_1\quad C \phi_2\quad C \phi_3\quad C \phi_4\quad C\phi_5\quad \cos\dfrac{\tilde{\phi}}{\sqrt{2} f} + \sqrt{\dfrac{3}{10}} C \phi_0 \Biggr)\,,\\[3mm]
C=\dfrac{i}{\tilde{\phi}} \sin \dfrac{\tilde{\phi}}{\sqrt{2} f}\,,\qquad \tilde{\phi}=\sqrt{\dfrac{3}{10}\phi_0^2+|\phi_1|^2+|\phi_2|^2+|\phi_3|^2+|\phi_4|^2+|\phi_5|^2}\,,
\end{array}
\label{quarks-3}
\end{equation}
where the $\mbox{SU(6)}$ generators are normalised so that $\mbox{Tr} T^a T^b = \frac{1}{2} \delta_{ab}$ and
$$
\Omega_0^T= (0\quad 0\quad 0\quad 0\quad 0\quad 1)\,,\qquad \Sigma= e^{i\Pi/f}\,,\qquad \Pi=\Pi^{\hat{a}} T^{\hat{a}}\,.
$$
In the leading approximation the Lagrangian, that describes the interactions of the pNGB states, can be written as
\begin{equation}
\mathcal{L}_{pNGB}=\frac{f^2}{2}\biggl|\mathcal{D}_{\mu} \Omega \biggr|^2\,.
\label{quarks-4}
\end{equation}
The field $\phi_0$ is real and does not participate in the $\mbox{SU(3)}_C\times \mbox{SU(2)}_W\times U(1)_Y$ gauge interactions. Five components of
vector $\Omega$, i.e $\tilde{H}\sim (\phi_1\,\, \phi_2\,\, \phi_3\,\, \phi_4\,\, \phi_5)$, form a fundamental representation of the unbroken $\mbox{SU(5)}$
subgroup of $\mbox{SU(6)}$. The components $H\sim (\phi_1\, \phi_2)$ transform as an $\mbox{SU(2)}_W$ doublet. Therefore $H$ corresponds to the SM--like
Higgs doublet. Three other components of $\tilde{H}$, i.e. $T\sim (\phi_3\,\, \phi_4\,\, \phi_5)$, transform as an $\mbox{SU(3)}_C$ triplet.
In the E$_6$CHM neither $H$ nor $T$ carry baryon and/or lepton number.

The pNGB effective potential $V_{eff}(\tilde{H}, T, \phi_0)$ is induced by the interactions of the elementary states with their composite partners,
which break $\mbox{SU(6)}$ global symmetry. The analysis of the structure of this potential including the derivation of quadratic terms $m^2_H |H|^2$
and $m_T^2 |T|^2$ in the composite Higgs models, which are similar to the E$_6$CHM, shows that there is a considerable part of the parameter
space where $m_H^2$ is negative and $m_T^2$ is positive \cite{Frigerio:2011zg}--\cite{Barnard:2014tla}. In this parameter
region the $\mbox{SU(2)}_W\times U(1)_Y$ gauge symmetry gets broken to $U(1)_{em}$, associated with electromagnetism, while $\mbox{SU(3)}_C$ colour
is preserved. Because in the E$_6$CHM the scale $f\gtrsim 10\,\mbox{TeV}$, a significant tuning, $\sim 0.01\%$, is required to get the appropriate value
of the parameter $m^2_H $ that results in a $125\,\mbox{GeV}$ Higgs state.

Since in the E$_6$CHM all states in the strongly interacting sector fill complete $\mbox{SU(5)}$ representations the corresponding fields contribute equally to
the beta functions of the $\mbox{SU(3)}_C$, $\mbox{SU(2)}_W$ and $U(1)_Y$ interactions in the one--loop approximation. As a consequence the convergence of the
SM gauge couplings is determined by the matter content of the weakly--coupled sector. In this case, approximate gauge coupling unification can be
achieved if the right--handed top quark, $t^c$, is entirely composite and the weakly--coupled elementary sector involves the following set of multiplets
(see also \cite{Agashe:2005vg}):
\begin{equation}
(q_i,\,d^c_i,\,\ell_i,\,e^c_i) + u^c_{\alpha} + \bar{q}+\bar{d^c}+\bar{\ell}+\bar{e^c}+\eta\,,
\label{quarks-5}
\end{equation}
where $\alpha=1,2$ runs over the first two generations and $i=1,2,3$ runs over all three.
In Eq.~(\ref{quarks-5}) $u_{\alpha}^c, d_i^c$ and $e_i^c$ represent the right-handed up- and down-type quarks and charged leptons,
$q_i$ and $\ell_i$ correspond to the left-handed quark and lepton doublets, whereas $\bar{q},\,\bar{d^c},\,\bar{\ell}$ and $\bar{e^c}$ are exotic states
which have opposite $\mbox{SU(3)}_C\times \mbox{SU(2)}_W\times U(1)_Y$ quantum numbers to the left-handed quark doublets, right-handed down-type quarks,
left-handed lepton doublets and right-handed charged leptons, respectively. An additional SM singlet exotic state, $\eta$, with spin $1/2$ is included
to ensure the phenomenological viability of the model under consideration. The set of elementary fermion states (\ref{quarks-5}) is chosen so that
the weakly--coupled sector contains all SM fermions except right--handed top quark and anomaly cancellation takes place.
Using the one--loop renormalisation group equations (RGEs) one can find the exact gauge coupling unification is attained for $\alpha(M_Z)=1/127.9$,
$\sin^2\theta_W=0.231$ and $\alpha_3(M_Z)\simeq 0.109$\,. The scale where the unification of the SM gauge couplings takes place
is somewhat close to $M_X\sim 10^{15}-10^{16}\, \mbox{GeV}$. This estimation demonstrates that for the phenomenologically acceptable values
$\alpha_3(M_Z)\simeq 0.118$ the SM gauge couplings can be reasonably close to each other at very high energies around
$M_X\simeq 10^{16}\, \mbox{GeV}$.

The scenario under consideration implies that the dynamics of the strongly coupled sector below the scale $f$ leads to the composite
${\bf 10} + {\bf \overline{5}} + {\bf 1}$ multiplets of $\mbox{SU(5)}$. Because of the conservation of the $U(1)_B$ and $U(1)_L$ charges
all components of the $10$--plet, i.e. $Q$, $E^c$ and $t^c$, carry the same baryon and lepton numbers as the right--handed top quark $t^c$.
The components of ${\bf \overline{5}}$ ($D^c$ and $L$) and ${\bf 1}$ ($\bar{\eta}$) can have baryon charges $-1/3$ and $+1/3$ \cite{e6chm}.
It is expected that the composite multiplets $Q$, $E^c$, $D^c$, $L$ and $\bar{\eta}$ get combined with the elementary exotic states $\bar{q}$,\,
$\bar{e^c}$,\,$\bar{d^c}$,\,$\bar{\ell}$ and $\eta$, respectively, giving rise to a set of vector--like fermion states. The only exceptions are
the components of the $10$--plet associated with the composite right--handed top quark which survive down to the electroweak scale.

In the E$_6$CHM the lightest exotic fermion state has to be stable. Indeed, the baryon number conservation implies that
the Lagrangian of the E$_6$CHM is also invariant under the transformations of the discrete $Z_3$ symmetry which can be defined as
\begin{equation}
\Psi \longrightarrow e^{2\pi i B_3/3} \Psi,\qquad B_3 = (3 B - n_C)_{\mbox{mod}\,\, 3}\,.
\label{quarks-6}
\end{equation}
Here $B$ is the baryon number of the given multiplet $\Psi$ and $n_C$ is the number of colour indices ($n_C=1$ for ${\bf 3}$
and $n_C=-1$ for ${\bf \overline{3}}$). This symmetry is called baryon triality \cite{Frigerio:2011zg}. All states in the SM
have $B_3=0$. At the same time exotic fermion states carry either $B_3=1$ or $B_3=2$. As a result the lightest exotic state
with non--zero $B_3$ can not decay into SM particles and should be stable. Since models with stable charged particles
are ruled out by various experiments \cite{quarks-1}-\cite{Hemmick:1989ns}, the lightest exotic fermion in the E$_6$CHM
must be neutral. It is also worth noting that the coupling of this neutral Dirac fermion to the $Z$--boson have to be extremely
suppressed. Otherwise this stable exotic state would scatter on nuclei resulting in unacceptably large spin--independent cross
sections. Thus, only a Dirac fermion, which is mostly a superposition of $\eta$ and $\bar{\eta}$, can be the lightest exotic state
in the E$_6$CHM.
\section{750 GeV diphoton resonance}
The SM singlet pNGB state $\phi_0$ can be identified with the 750 GeV diphoton resonance recently reported by ATLAS and CMS.
It is important that no $750\,\mbox{GeV}$ resonance has been observed in other channels like $pp\to t\bar{t}, WW, ZZ, b \bar{b},
\tau\bar{\tau}$ and $jj$. This may be an indication that the detected signal is just a statistical fluctuation. At the same time, if these
observations are confirmed this should set stringent constraints on the new physics models that may lead to such a signature.
For example, in the E$_6$CHM the field $\phi_0$ can mix with the Higgs boson which would result in large partial widths of
the $750\,\mbox{GeV}$ resonance associated with the decays of this state into pairs of $Z$-bosons, $W$-bosons and $t\bar{t}$.
The corresponding mixing can be suppressed if invariance under the CP transformation is imposed. Indeed, in this case $\phi_0$
manifests itself in the Yukawa interactions with fermions as a pseudoscalar field. In particular, the couplings of the SM singlet
pNGB state $\phi_0 =A$ to the top quarks is induced by
\begin{equation}
\mathcal{L}_{AT}= \frac{y_t}{\Lambda_t} A (i\bar{t}_L H^0 t_R + h.c.)\,.
\label{quarks-7}
\end{equation}
Because of the almost exact CP--conservation the mixing between the Higgs boson
and pseudoscalar $A$ is forbidden.

The Lagrangian that describes the interactions between $A$ and exotic fermions can be written in the following form
\cite{Nevzorov:2016fxp}
\begin{equation}
\mathcal{L}_{AF}= A \Biggl(i\kappa_D  \bar{d^c} D^c + i \kappa_Q \bar{q} Q + i \lambda_L  \bar{\ell} L +
i \lambda_E  \bar{e^c} E^c + i\lambda_{\eta} \bar{\eta}\eta +h. c. \Biggr)\,.
\label{quarks-8}
\end{equation}
In the most general case the couplings $\kappa_i$ and $\lambda_i$ in Eq.~(\ref{quarks-8}) and the exotic fermion masses $\mu_i$,
induced below scale $f$, i.e.
\begin{equation}
\mathcal{L}_{mass}= \mu_D \bar{d^c} D^c + \mu_Q \bar{q} Q + \mu_L  \bar{\ell} L +
\mu_E  \bar{e^c} E^c + \mu_{\eta} \bar{\eta}\eta +h.c.\,,
\label{quarks-9}
\end{equation}
are entirely independent parameters, which are not constrained by the $\mbox{SU(6)}$ and $\mbox{SU(5)}$ symmetries.
In order to get the cross section $\sigma(pp\to\gamma\gamma)$, which corresponds to the production and
sequential diphoton decays of the pseudoscalar $A$, of about $5-10\,\mbox{fb}$
we assume that $\mu_D$,\,$\mu_Q$,\, $\mu_L$, $\mu_E$ and $\mu_{\eta}$ are larger than
$375\,\mbox{GeV}$. As a result the on-shell decays of $A$ into the exotic fermions, that result in the strong
suppression of the branching ratios of the decays of this pNGB state into photons, are not kinematically allowed.
Integrating out the exotic fermion states one obtains the effective Lagrangian
that describes the interactions of the pseudoscalar $A$ with the SM gauge bosons \cite{Nevzorov:2016fxp}
\begin{equation}
\mathcal{L}^A_{eff} = c_1 A B_{\mu\nu} \widetilde{B}^{\mu\nu} + c_2 A W^a_{\mu\nu} \widetilde{W}^{a\mu\nu} +
c_3 A G^{\sigma}_{\mu\nu} \widetilde{G}^{\sigma\mu\nu}\,,
\label{quarks-10}
\end{equation}
where $B_{\mu\nu}$, $W^a_{\mu\nu}$, $G^{\sigma}_{\mu\nu}$ are field strengths for the $U(1)_Y$, $\mbox{SU(2)}_W$ and $\mbox{SU(3)}_C$
gauge interactions, $\widetilde{G}^{\sigma\mu\nu}=\frac{1}{2}\epsilon^{\mu\nu\lambda\rho} G^{\sigma}_{\lambda\rho}$,
$\widetilde{W}^{a\mu\nu}=\frac{1}{2}\epsilon^{\mu\nu\lambda\rho} W^{a}_{\lambda\rho}$,
$\widetilde{B}^{\mu\nu}=\frac{1}{2}\epsilon^{\mu\nu\lambda\rho} B_{\lambda\rho}$, whereas
\begin{equation}
\begin{array}{rcl}
c_1&=&\frac{\alpha_Y}{16\pi}\Biggl[\frac{2\kappa_{D}}{3\mu_{D}} B(x_{D}) + \frac{\kappa_{Q}}{3\mu_{Q}} B(x_{Q})
+  \frac{\lambda_{L}}{\mu_{L}} B(x_{L}) + 2 \frac{\lambda_{E}}{\mu_{E}} B(x_{E}) \Biggr]\,,\\[0mm]
c_2&=&\frac{\alpha_2}{16\pi}\Biggl[3 \frac{\kappa_{Q}}{\mu_{Q}} B(x_{Q}) +  \frac{\lambda_{L}}{\mu_{L}} B(x_{L}) \Biggr]\,,\\[0mm]
c_3&=&\frac{\alpha_3}{16\pi}\Biggl[\frac{\kappa_{D}}{\mu_{D}} B(x_{D}) + 2 \frac{\kappa_{Q}}{\mu_{Q}} B(x_{Q})\Biggr]\,,\\[0mm]
B(x) &=& 2 x \arcsin^2[1/\sqrt{x}]\,,\qquad \mbox{for} \qquad x\ge 1\,.
\end{array}
\label{quarks-11}
\end{equation}
In Eq. (\ref{quarks-11}) $x_{D}=4\mu_{D}^2/m_A^2$, $x_{Q}=4\mu_{Q}^2/m_A^2$, $x_{L}=4\mu_{L}^2/m_A^2$, $x_{E}=4\mu_{E}^2/m_A^2$,
$m_A \simeq 750\,\mbox{GeV}$ is the mass of the SM singlet pNGB state $A$, $\alpha_Y=3\alpha_1/5$ while $\alpha_1$, $\alpha_2$ and $\alpha_3$
are (GUT normalised) gauge couplings of $U(1)_Y$, $\mbox{SU(2)}_W$ and $\mbox{SU(3)}_C$ interactions.
Using Eqs.~(\ref{quarks-10})--(\ref{quarks-11}) one can obtain an analytical expression for the coupling of the pseudoscalar $A$
to the electromagnetic field $F_{\mu\nu}$
\begin{equation}
\mathcal{L}^{A\gamma\gamma}_{eff} = c_{\gamma} A F_{\mu\nu} \widetilde{F}^{\mu\nu}\,,\qquad
c_{\gamma} = c_1 \cos^2\theta_W + c_2 \sin^2\theta_W\,,
\label{quarks-12}
\end{equation}
where $\theta_W$ is the weak mixing (Weinberg) angle and $\widetilde{F}^{\mu\nu}=\frac{1}{2}\epsilon^{\mu\nu\lambda\rho} F_{\lambda\rho}$.

Since at the LHC the pseudoscalar $A$ is predominantly produced through gluon fusion the cross section
$\sigma_{\gamma\gamma}=\sigma(pp\to A\to \gamma\gamma)$ can be presented in the
following form \cite{Franceschini:2015kwy}
\begin{equation}
\sigma_{\gamma\gamma} \simeq \frac{C_{gg}}{m_A s} \frac{\Gamma(A \to gg) \Gamma (A\to \gamma\gamma)}{\Gamma_A}
\simeq 7.3\, \mbox{fb}\times \left(\frac{\Gamma(A\to gg) \, \Gamma(A \to \gamma\gamma)}{\Gamma_A \, m_A} \times 10^{6}\right)\,,
\label{quarks-13}
\end{equation}
where $C_{gg}\simeq 3163$, $\sqrt{s}\simeq 13\,\mbox{TeV}$, $\Gamma_A$ is a total width of the pseudoscalar $A$
while partial decay widths $\Gamma(A \to \gamma\gamma)$ and $\Gamma(A \to gg)$ are given by
\begin{equation}
\Gamma(A \to gg)=\frac{2 m_A^3}{\pi} |c_3|^2\,, \qquad
\Gamma(A \to \gamma\gamma)=\frac{m_A^3}{4 \pi} |c_{\gamma}|^2\,.
\label{quarks-14}
\end{equation}

\begin{figure}[h]
\hspace*{-24mm}{$\sigma(pp\to A \to \gamma\gamma)[\mbox{fb}]$}\\[-6mm]
\centering
\includegraphics[width=10cm,clip]{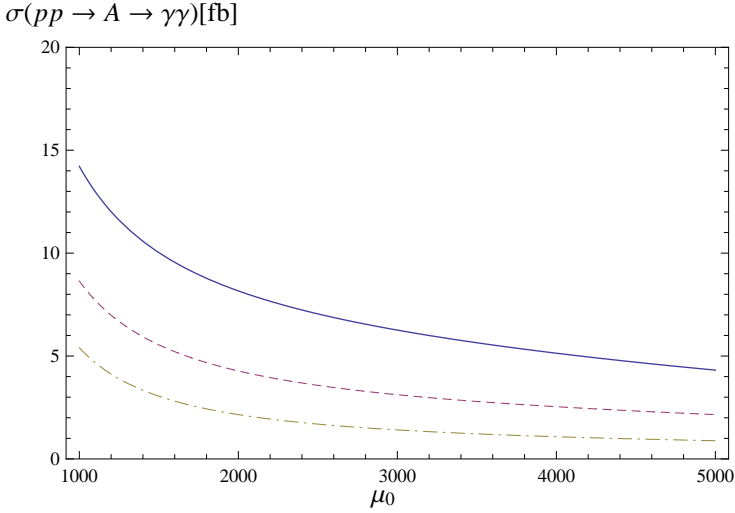}\\[-7mm]
\hspace*{0cm}{$\mu_0$}
\caption{The cross section of $\sigma(pp\to A \to \gamma\gamma)$ is shown as a function of
$\mu_Q=\mu_D=\mu_L=\mu_0$ for $\mu_E=400\,\mbox{GeV}$ (solid lines), $\mu_E=500\,\mbox{GeV}$ (dashed lines)
and $\mu_E=800\,\mbox{GeV}$ (dashed--dotted lines), for the case $\Lambda_t=80\,\mbox{TeV}$ and
$\kappa_D = \kappa_Q = \lambda_L = \lambda_E = \sigma=1.5$.}
\label{fig-quarks-1}
\end{figure}

First of all it is worthwhile to identify the scenario that leads to the suppression of the decay rates $A \to t\bar{t}, WW, ZZ, \gamma Z$
because no indication of the $750\, \mbox{GeV}$ resonance has been observed in the channels associated with
these decay modes. The analytical expressions for the corresponding partial decay widths can be presented in the following form
\begin{equation}
\Gamma(A \to t\bar{t})=\frac{3 m_A m_t^2}{8\pi \Lambda_t^2}\sqrt{1-\frac{4 m_t^2}{m_A^2}},\qquad\qquad\qquad\qquad
\label{quarks-15}
\end{equation}
\begin{equation}
\Gamma(A \to W W)=\frac{m_A^3}{2 \pi} |c_2|^2 \left(1-\frac{4 M_W^2}{m_A^2}\right)^{3/2},\qquad\qquad\qquad\qquad
\label{quarks-16}
\end{equation}
\begin{equation}
\Gamma(A \to ZZ)=\frac{m_A^3}{4 \pi} \Biggl|c_1 \sin^2 \theta_W + c_2 \cos^2 \theta_W \Biggr|^2 \left(1-\frac{4 M_Z^2}{m_A^2}\right)^{3/2}\,,\\[3mm]
\label{quarks-17}
\end{equation}
\begin{equation}
\Gamma(A \to \gamma Z)= \frac{m_A^3}{8 \pi} \sin^2 2\theta_W |c_1-c_2|^2 \left(1-\frac{M_Z^2}{m_A^2}\right)^3\,.\qquad\qquad
\label{quarks-18}
\end{equation}
In the model under consideration $\Lambda_t\simeq \sqrt{15} f$ if $t^c$ is mainly a component of ${\bf 20}^t$ of $\mbox{SU(6)}$.
Since $f\gtrsim 10\,\mbox{TeV}$ the partial decay width (\ref{quarks-15}) tends to be rather small, i.e.
$\Gamma(A \to t\bar{t})\lesssim \Gamma(A \to \gamma\gamma)$. Here we set $\Lambda_t\simeq 80\,\mbox{TeV}$.
The partial decay widths (\ref{quarks-16})--(\ref{quarks-18}) become substantially
smaller than $\Gamma(A \to \gamma\gamma)$ if $|c_2|\ll |c_1|$. The appropriate suppression of $|c_2|$ can be achieved
when the exotic fermions that form $\mbox{SU(2)}_W$ doublets are considerably heavier than the $\mbox{SU(2)}_W$ singlet exotic states.
On the other hand the non-observation of any new coloured particles with masses below $1\,\mbox{TeV}$ at the LHC
implies that the exotic coloured fermions in the E$_6$CHM should be rather heavy. Thus to simplify our numerical analysis
we assume that $\mu_D = \mu_Q = \mu_L = \mu_0 \gtrsim \mu_E$ and $\kappa_D = \kappa_Q = \lambda_L = \lambda_E = \sigma$.
For $\mu_0\gg \mu_E$ the decay rates for $A \to t\bar{t}, WW, ZZ$ and $Z\gamma$ are very suppressed. However
$\mu_0$ cannot be too large, otherwise the LHC production cross section of the pseudoscalar $A$ becomes smaller
than $5-10\,\mbox{fb}$. In our numerical analysis we vary $\mu_0$ from $1\,\mbox{TeV}$ to $5\,\mbox{TeV}$.
In this case $A$ mainly decays into a pair of gluons. As a consequence $\Gamma_A\approx \Gamma(A \to gg)$ and the cross
section (\ref{quarks-13}) is determined by $\Gamma(A \to \gamma\gamma)$. Then the value of $\mu_E$ can be adjusted so
that $(\Gamma(A \to \gamma\gamma)/m_A) \sim 10^{-6}$ leading to $\sigma(pp\to A \to \gamma\gamma)\simeq 5-10\,\mbox{fb}$.

\begin{figure}[h]
\hspace*{-10mm}{$\mbox{BR}(A \to t\bar{t}, gg, \gamma\gamma, WW, ZZ, \gamma Z)$}\\[-6mm]
\centering
\includegraphics[width=10cm,clip]{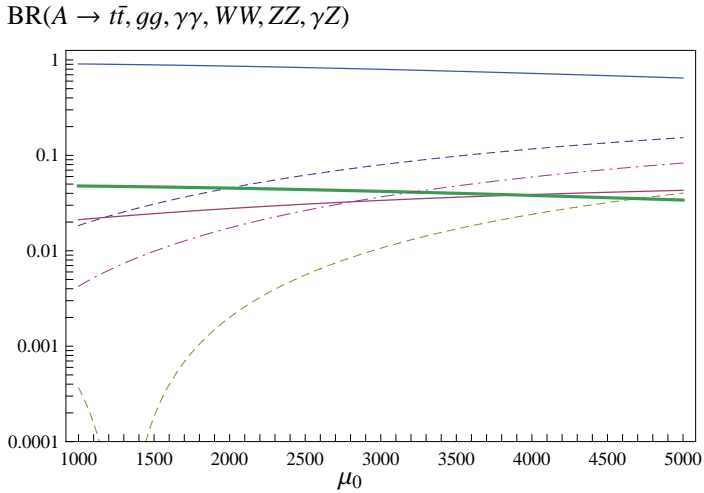}\\[-7mm]
\hspace*{0cm}{$\mu_0$}
\caption{The branching ratios of the decays of the pseudoscalar $A$ into $t\bar{t}$ (dashed--dotted lines), $gg$ (highest solid lines),
$\gamma\gamma$ (highest dashed lines), $WW$ (thick solid lines), $ZZ$ (lowest solid lines) and $\gamma Z$ (lowest dashed lines)
are presented as a function of $\mu_Q=\mu_D=\mu_L=\mu_0$ for $\mu_E=400\,\mbox{GeV}$, $\Lambda_t=80\,\mbox{TeV}$ and
$\kappa_D = \kappa_Q = \lambda_L = \lambda_E = \sigma=1.5$.}
\label{fig-quarks-2}
\end{figure}

Fig.~1 demonstrates that $\sigma(pp \to A \to \gamma\gamma)$ decreases very substantially when $\mu_{E}$ increases.
For $\mu_E=400\,\mbox{GeV}$ and $\sigma\gtrsim 1.5$ the cross section (\ref{quarks-16}) can be about of $5\,\mbox{fb}$,
even when all other exotic fermions are rather heavy $\mu_0\simeq 5\,\mbox{TeV}$. At the same time for $\mu_E\simeq 700-800\,\mbox{GeV}$
the corresponding diphoton production cross section becomes sufficiently large only if $\mu_0\simeq 1\,\mbox{TeV}$.
When $\mu_0$ changes from $5\,\mbox{TeV}$ to $1\,\mbox{TeV}$ the ratio $\Gamma_A/m_A$ increases from $10^{-5}$ to
$10^{-4}$ that corresponds to the variation of the total LHC production cross section of the pseudoscalar $A$ from
$100\,\mbox{fb}$ to $1\,\mbox{pb}$.

In Fig.~2 the dependence of the branching ratios of the pseudoscalar $A$ on $\mu_0$ is explored
for $\mu_E\simeq 400\,\mbox{GeV}$ and $\sigma=1.5$. From this figure it follows that $A\to gg$ is the dominant decay channel.
Its branching fraction is always close to $100\%$. When $\mu_0\simeq 5\,\mbox{TeV}$ the branching ratio $\mbox{BR}(A \to \gamma\gamma)$
is the second largest one. $\mbox{BR}(A \to W W)$, $\mbox{BR}(A \to Z Z)$, $\mbox{BR}(A \to Z\gamma)$ and $\mbox{BR}(A \to t\bar{t})$ are
substantially smaller than $\mbox{BR}(A \to \gamma\gamma)$. This might be a reason why the decays $A \to W W,\,ZZ,\, Z\gamma, t\bar{t}$ have
not been detected yet. The decays $A\to gg$ can be rather problematic to observe because the total LHC production cross section of the
pseudoscalar $A$ is quite small. The branching fractions $\mbox{BR}(A \to Z Z)$ and $\mbox{BR}(A \to WW)$ increase while
$\mbox{BR}(A \to \gamma\gamma)$ and $\mbox{BR}(A \to t\bar{t})$ decrease with decreasing $\mu_0$. When $\mu_0\simeq 1\,\mbox{TeV}$
the branching ratios $\mbox{BR}(A \to Z Z)$ and $\mbox{BR}(A \to WW)$ are somewhat bigger than $\mbox{BR}(A \to \gamma\gamma)$.
Nevertheless the experimental detection of $A \to Z Z$ and $A \to WW$ can be rather difficult since the $W$ and $Z$ bosons decay
predominantly into quarks. In this scenario $\mbox{BR}(A \to \gamma Z)$ remains the lowest branching fraction and vanishes at some
value of $\mu_0$.

\begin{acknowledgement}
This work was supported by the University of Adelaide and the Australian Research Council through
the ARC Center of Excellence in Particle Physics at the Terascale (CE 110001004) and through grant LF099 2247 (AWT).
\end{acknowledgement}

\end{document}